\documentclass[a4paper,fleqn,usenatbib]{mnras}

\usepackage{newtxtext,newtxmath}

\usepackage[T1]{fontenc}
\usepackage{ae,aecompl}

\usepackage{graphicx}	% Including figure files
\usepackage{amsmath}	% Advanced maths commands
\usepackage{amssymb}	% Extra maths symbols

\title[Galaxies and Clusters of Galaxies as Peak Patches of the Density Field]{
Galaxies and Clusters of Galaxies as Peak Patches of the Density Field}

\author[Masataka Fukugita]{
Masataka Fukugita,$^{1,2}$\thanks{E-mail:  fukugita@icrr.u-tokyo.ac.jp}
Hans B\"ohringer$^{3,4}$
\\
$^{1}$Kavli Institute for the Physics and Mathematics of the Universe,
                   University of Tokyo, Kashiwa 2778583, Japan\\
$^{2}$Institute for Advanced Study, Princeton, NJ08540, USA\\
$^{3}$Universit\"atssternwarte M\"unchen, Ludwig-Maximilians-Universit\"at,
                   Scheinerstr. 1, D-81679,  M\"unchen, Germany\\
$^{4}$Max-Planck-Institut f\"ur extraterrestrische Physik,
                   D-85748 Garching, Germany.
}

\date{Accepted XXX. Received YYY; in original form ZZZ}

\pubyear{2015}

\begin{document}
\label{firstpage}
\pagerange{\pageref{firstpage}--\pageref{lastpage}}
\maketitle

\begin{abstract}
The mass function of galaxies and clusters of galaxies
can be derived observationally based on different types of observations.
In this study we test if these observations can be combined to a consistent
picture which is also in accord with structure formation theory. The galaxy data
comprise the optical galaxy luminosity function and the gravitational lensing 
signature of the galaxies, while the galaxy cluster mass function is derived from
the X-ray luminosity distribution of the clusters. We show the
results of the comparison in the form of the mass density fraction that is contained
in collapsed objects relative to the mean matter density in the Universe.
%We find that
The mass density fraction in groups and clusters of galaxies
extrapolated to low masses agrees
very well with that of the galaxies: both converge at the low mass limit
to a mass fraction of about 28\% if the outer radii of the objects are taken to
be $r_{200}$. Most of the matter contained in collapsed objects is found in 
the mass range $M_{200} \sim  10^{12} - 10^{14} h^{-1}_{70} M_\odot$,
while a larger amount of the cosmic matter resides outside of 
$r_{200}$ of collapsed objects.
\end{abstract}


%%%%%%%%%%%%%%%%%%%%%%%%%%%%%%%%%%%%%%%%%%%%%%%%%%

%%%%%%%%%%%%%%%%% BODY OF PAPER %%%%%%%%%%%%%%%%%%

\section{Introduction}

In modern theory of cosmological structure formation, it is
supposed that galaxies and clusters of galaxies formed from peak
patches of the density field of matter in the Universe (Bardeen et
al. 1986). In cosmological simulations the primary reference objects 
which are populated by galaxies and galaxy clusters are dark matter 
halos and their abundance is described by the dark matter halo 
mass function (e.g. Jenkins et al. 2001, Tinker et al. 2008). 
Observationally galaxies and glaxy clusters have very different
appearances. Galaxies just mark the central region of the 
dark matter halo and the extent of their embedding
dark matter halo can only be traced by weak gravitational lensing.
On the contrary the dark matter halos of clusters of galaxies are
filled by a hot, X-ray luminous intracluster plasma, 
which can  easily be observed with X-ray telescopes
(e.g. Sarazin, 1986) and through the Sunyaev-Zeldovich 
effect in the cosmic microwave background (Sunyaev \& Zeldovich, 1972). 
In this way the gravitational potential of the 
dark and baryonic matter halo can be visualised more directly.

In this note we explore if the observational data on galaxies and
groups and clusters of galaxies can be described consistently in the 
from of a continous halo mass function, even though the observational
signatures of these objects are very different. We test in this way 
the validity of structure formation theory and the correctenss 
of the interpretation of the observational data. In the present study
we show as representation of the object mass
distribution mostly the fraction of the cosmic matter density made
up by galaxies and clusters, which is a direct reflection of the
cumulative mass function. This provides us in addition with the
interesting information where the major parts of matter are 
located in our Universe.

For all calculations depending on the cosmological model, we use
a flat cosmic geometry and the parameters, $\Omega_m = 0.282$
\footnote{The uncertainty of $\Omega_m$ is degenerate with that of
$\sigma_8$ and can be represented as $\sigma_8 (\Omega_m/0.3)^{0.57} 
= 0.75 \pm 0.03$ (B\"ohringer et al. 2014).}
(B\"ohringer et al. 2017) and $H_0 = 70$ km s$^{-1}$ Mpc$^{-1}$ . 
We retain $h = h_{100}$ for some values quoted from the literature.
This mass density is compared
to $0.308\pm0.012$ of the 2015 result of Planck (Planck Collaboration
2016) and to $0.279\pm0.025$ of the WMAP 9 year result (Bennett et al. 2013). 
The specific value for 
$\Omega_m$ is chosen because it provides the best fit to our
data on the galaxy cluster abundance and we thus apply it in
the following for consistency reasons.
The deviation from the Planck result agrees with the
well-recognised tension seen in the $\sigma_8-\Omega_m$ plane 
between the Planck result and that from
weak lensing; see e.g. Hildebrandt et al. (2017). The cluster fit
gives a value consistent with the weak lensing result. 

\section{Galaxy and cluster data}

To assign a definte mass to galaxies and their dark matter halos and
to galaxy clusters, we need to define an outer radius up to
which the mass distribution in the systems is integrated.
In an analysis of gravitational lensing around galaxies it is
indicated that the mass of galaxies is distributed beyond the
pseudovirial radius of galaxies, which was operationally
defined as the radius, $r_{200}$, which
encircles a mass corresponding to 200 times the critical density
(Masaki et al. 2012; hereafter MFY).  
The analysis indicates that the distribution of mass around
galaxies is extended to a few Mpc, to the middle to neighbouring
galaxies: there seems no boundary in the mass distribution.
Also for galaxy clusters the mass profile continues to increase
well beyond a radius of $r_{200}$; see e.g. Ettori et al. (2019). 
Since there is no clear, natural outer edge to these collapsed 
objects, a common fiducial outer radius has to be adopted for the comparison of
the galaxy and cluster matter density content. Here we use $r_{200}$
in our further analysis, which approximately describes the boundary
between the partly virialised material inside and the mostly infalling
matter outside.

\begin{figure}
	\includegraphics[width=\columnwidth]{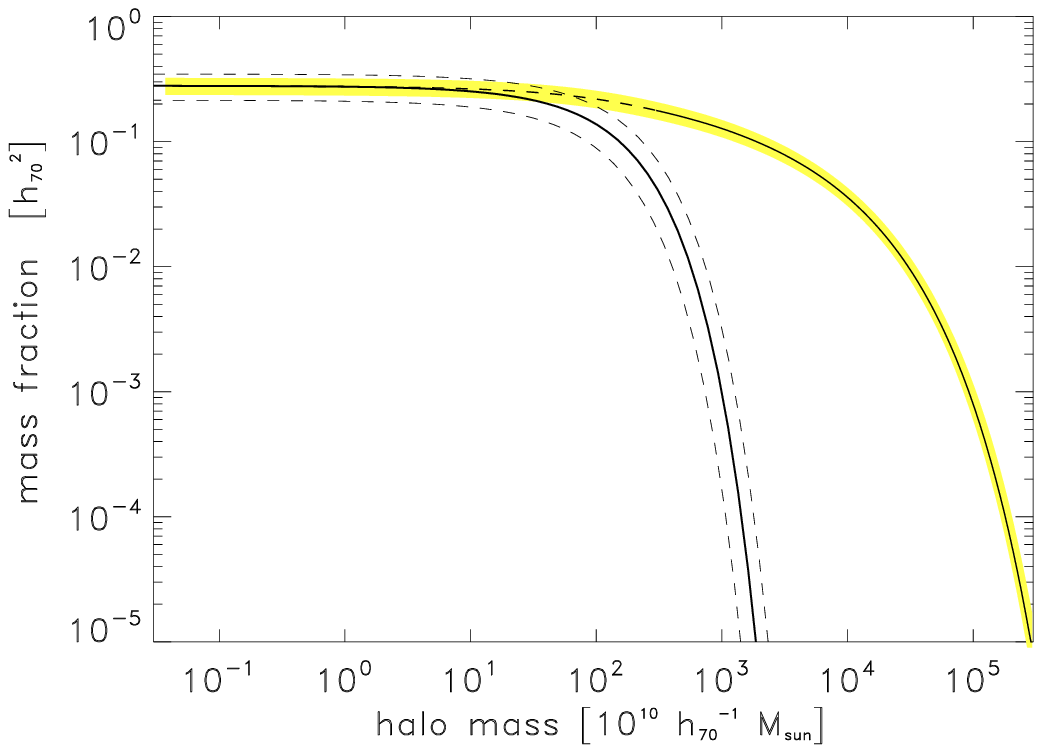}
    \caption{Fraction of the matter of the Universe contained in collapsed
objects (inside $M_{200}$) with masses above the lower mass limit given on the x-axis. 
The curve on the right gives the mass fraction in groups and
clusters of galaxies, where the solid line marks the function derived
from observations with uncertainties given by the grey area. The extrapolation
to lower masses, by means of the halo mass function of Tinker et al.
(2008), is indicated by a dashed line. The curve to the left is the
fraction of the matter density contained in galaxy halos deduced from
the galaxy luminosity function and the gravitational lensing effect of
galaxy halos. The region of the curve in which the data are extrapolated from 
the interval covered by observations is shown as dashed line.  The grey shaded
area and the thin dashed lines indicate the uncertainty ranges of the functions.
}
    \label{fig1}
\end{figure}

The mass function of galaxy halos for our study  was obtained in the 
following way. The luminosity function of galaxies is now accurately
known (Blanton et al. 2001; 2003; Folkes 1999) to $L>10^{8.3}L_\odot$. 
Here we use the Blanton et al. 2001 luminosity function,
  which refers to the standard $ugriz$ photometric system\footnote{
    We note a discrepancy in the luminosity density between their 2001 and
    2003 luminosity functions, giving rise to the luminosity density which
    is 30\%
    smaller in their 2003 publication.
}.
McKay et
al. (2001; 2002) measured the mass of galaxies encircled by haloes to
$260h^{-1}$kpc by measuring weak lensing shear around galaxies for the
Sloan Digital Sky Survey (SDSS) spectroscopic sample.  Their
measurement gives $\langle M/L_r\rangle \simeq 170\pm21h^{-1}$ for the
$r$-band for the mass of galaxies encircled by haloes to $260h^{-1}$kpc,
which is thought to be well beyond the virial radius of galaxies and
thus to stand for the mass associated with galaxies. Their data show
that the mass-to-light ratio does not depend on galaxy luminosity 
for an interval of a decade, $5 \times 10^9 - 8 \times 10^{10}$ L$_{\odot}$.  
They also find the dynamical mass from
the virial velocity for the same sample to be $\langle
M/L_r\rangle \simeq 145\pm34h^{-1}$, with a reasonable agreement with
their lensing estimate.  For our analysis we adopted $160\pm30h^{-1}$
at the radius of $260h^{-1}$kpc, but scaled to the pseudovirial radius.

With the aid of the N-body simulation result for
haloes of galaxies, the average pseudovirial radius 
($r_{200}$) of galaxies that match
the SDSS sample, which is estimated to have a lower mass cutoff
$M_{\rm low}\approx 2\times 10^{11}h^{-1}M_\odot$, is approximately 
$120h^{-1}$kpc, and so the radius McKay et al. measured corresponds to
$\approx 2.2r_{200}$ (MFY).  As $260h^{-1}$kpc is significantly
larger than $r_{200}$, this is taken as evidence that the mass distribution
extends much beyond $r_{200}$; $r_{200}$ comprise only a fraction of mass
associated with galaxies.  For the comparison with clusters, we scale 
the average mass measured at $260h^{-1}$kpc to that at $r_{200}$,
using the weak lensing scaling
result, which approximately reads $M\propto r^{0.6}$ beyond the
pseudovirial radius (MFY).
This yields $\langle M/L_r\rangle|_{r_{200}} \simeq 90\pm 20h^{-1}$.  
This is the value we have adopted  
to estimate the mass of galaxies. 

We remark that this radius dependence of the mass profile 
is consistent with that expected
for the Navarro-Frenk-White (NFW, Navarro et al. 1995, 1997) 
profile with the core radius $r_s$
in units of $r_{200}$ to be $c=r_{200}/r_s=5-10$, which is the value
compatible to that derived for clusters $c\approx 5$ and for haloes of
galaxies $c\approx 10-15$ from inner profiles, typically, for
$r<r_{500}$. This means that the NFW profile stands also for a good
description of galaxy haloes extended beyond the virial radius.
Combing the galaxy luminosity function with the mass-to-light 
ratio from weak lensing we construct the galaxy halo mass function. 

In our preceding work (B\"ohringer et al. 2017) we have computed the
mass function of clusters and groups of galaxies down to
$3\times10^{12}h^{-1}_{70}M_\odot$, using an X-ray selected
cluster-group sample. We find that this mass function agrees well with
that obtained from optical cluster samples (Bahcall \& Cen 1993), when
the cluster mass is standardised to a universal definition, say by
adopting $r_{200}$. 

The mass function of groups and clusters in (B\"ohringer et al. 2017)
was derived from the cluster catalogue compiled in the {\sf REFLEX II} survey 
which was based on X-ray detections of clusters in the ROSAT All Sky Survey
in the southern sky (B\"ohringer et al. 2013). Since X-ray luminosity is tightly
correlated with the cluster mass, the X-ray selection of the galaxy
clusters is a good basis for the construction of the cluster mass function.
The cluster sample fulfils another important requirement being statistically highly
complete (~95\%) and described by a well understood selection function. The
cluster catalog is flux-limited with a minimum unabsorbed X-ray flux of 
$1.8 \times 10^{-12}$ erg s$^{-1}$ cm$^{-2}$ in the 0.1 - 2.4 keV 
energy band (B\"ohringer et al. 2004, 2013). 
The cluster sample used covers the redshift range $z = 0 - 0.4$ and has
a median redshift of $z = 0.1$, very similar to the SDSS galaxy sample.
The important observational census on which 
the further work is based is the X-ray luminosity function 
(B\"ohringer et al. 2014).

The mass function was determined from the luminosity function in
two different ways. Cluster masses were estimated by means of the 
X-ray luminosity -- mass relation determined for smaller subsamples 
(Vikhlinin et al. 2009, Pratt et al. 2009). By this means the 
X-ray luminosity function was converted into the cluster mass
function. In the second approach to constrain the cluster mass function
we use our observational data to fit them to cosmological model predictions
for the X-ray luminosity function of clusters. This fit was used in
B\"ohringer et al. (2014) to constrain cosmological model parameters. 
The theoretical prediction of the cluster X-ray luminosity function
for a given set of cosmological parameters involved the following steps:
we adopt a $\Lambda$CDM cosmological model with flat geometry,
the matter density distribution power spectrum was determined 
with CAMB (Lewis et al. 2000)\footnote{ CAMB is publicly available from
  http://www.camb.info/CAMBsubmit.html}.
We take a parametrized form of the halo mass 
function derived from N-body simulations by Tinker et al. (2008)
to construct the prediction for the cluster mass function.
The empirical cluster mass - X-ray luminosity relation with its
scatter and uncertainties is used to finally compare to the
observed X-ray luminosity function.
The statitical uncertainty of the most critical
cosmological parameters, $\Omega_m$ and $\sigma_8$
(B\"ohringer et al. 2017, Fig. 1) , and the errors on the
$L_x$ - $M$ scaling relation in the fit determins
the uncertainty range of the mass function. The two ways to obtain
the mass function are in good agreement, as shown in B\"ohringer et al. 
(2017, Fig. 2).

For the present work we use the constraints on the cluster mass function 
from the method that involves the fit to the cosmological model predictions
for two reasons: this method provides tighter constraints since it
includes our knowledge about cosmic structure formation,
and second the theoretical framework allows us
to extrapolate the mass function beyond the observational limits.
The observational data 
of the cluster sample cover the mass range 
$M_{200} = 7\times10^{12}$ to $3\times10^{15} h^{-1}_{70}M_\odot$. 
In the present work we add the uncertainty of the numerically 
derived mass function, which is in the range of 
5 - 10\%, (Tinker et al. 2008) as an additional uncertainty of
conservatively 10\% to the resulting mass function. For the comparison with
the galaxy data we use the mass function derived for a redshift of
$z = 0.1$, which is also the fiducial redshift for the galaxy sample.
We compare
these results to the mass function obtained using other parametrisations
for the halo mass function from the literature (e.g. Watson et a. 2013,
Despali et al. 2016), finding that differences lie well within our uncertainty.

\section{The combined matter density fraction}

From the galaxy and cluster mass function determined as described above,
we derive the matter fraction contained in all collapsed objects
inside $M_{200}$ 
above a certain limiting mass. For these calculations we have taken
$\Omega_m = 0.282$ consistent with the best fit to the cluster abundance
(B\"ohringer et al. 2017).
The matter fraction was calculated from $\rho_m^{-1}~ \int {dn/dm} ~m ~dm$,
where ${dn/dm}$ is the differential cluster mass function.  
Fig. 1 shows the mass fraction in collapsed objects from galaxies to groups
and clusters of galaxies. The dashed part of the cluster mass function
shows the regime where the mass function is extrapolated to
masses lower than covered by the observational data.
The galaxy halo mass fraction was estimated from the luminosity function
of galaxies (Blanton et al. 2001) multiplied with the mass-to-light ratio,
$\rho_m={\cal L}_r\times \langle M/L_r\rangle$,
where  $\langle M/L_r\rangle \simeq 90\pm 20h^{-1}$
and $\cal L_{\rm r}$ is the galaxy luminosity density in the $r$ band. 
The galaxy halo mass function is observationally constraint 
to $M>10^{11.2}M_\odot$. We note that at
the low masses the two functions match perfectly, even though they
have been derived from very different observational data sets
\footnote{The luminosity function is still uncertain and this
  nearly perfect match would be disturbed up to 30\% if we adopt
  Blanton et al 2003. The resuls, however, would still be consistent
  within the combined error limits.}.

In Fig. 2 we show the differential form of the matter fraction derived
from galaxy group and cluster observations. It is derived from the mass function 
through $\rho_m^{-1} {dn \over d\ln ~m} ~m $, giving the mass fraction
per ${\rm ln}~ m$ interval.
%, where we have used the natural logarithm.
This curve illustrates, which object population
contributes most to the matter density. We see a broad maximum for the 
mass range $M_{200} \sim 10^{12} - 10^{14}~ h^{-1}_{70} M_\odot$.

\begin{figure}
   \includegraphics[width=\columnwidth]{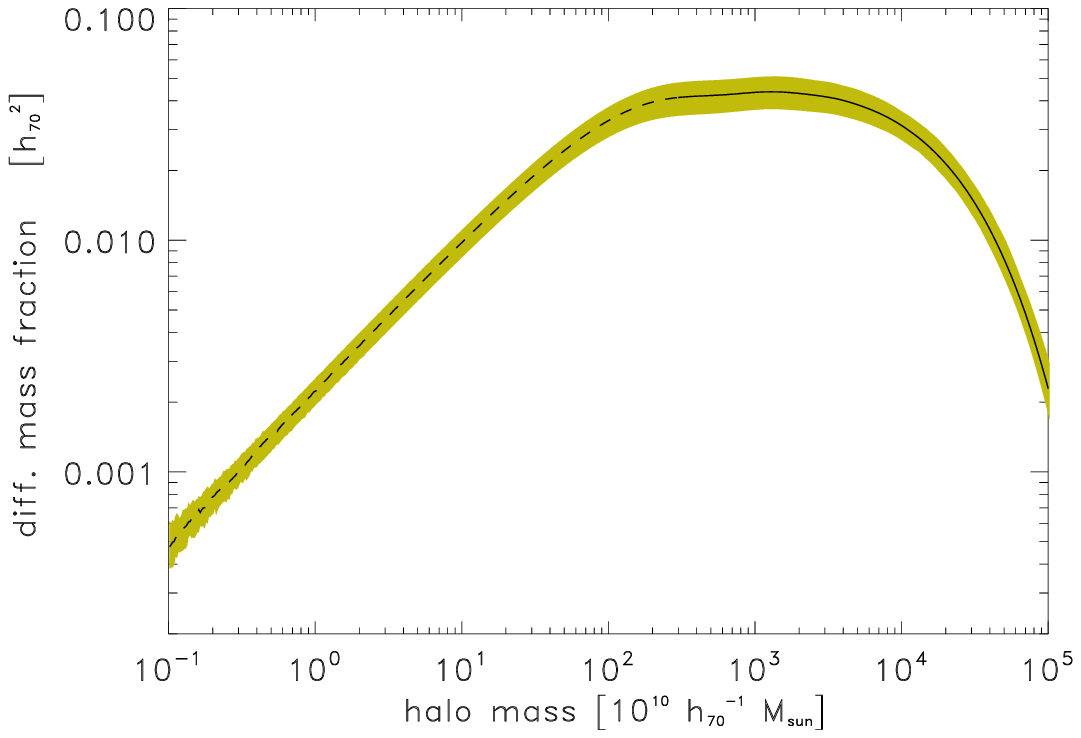}
\caption{Differential matter fraction for groups and 
clusters of galaxies. The function gives the mass 
fraction per $ln~ m$ interval.
The grey shaded area indicates the uncertainty
of the function.
}
\label{fig2}
\end{figure}

Fig. 3 shows the local power law index (logarithmic slope) of the 
cumulative mass function and of the function of the matter fraction of 
groups and clusters of galaxies. We find that the matter fraction saturates
at masses lower than about $10^{11} M_\odot$, with a further increase of 
not more than 1\%. This originates from a flattening of the cumulative
mass function. In our previous study we have fitted a Schechter function
as an approximation to the observed cumulative mass function of groups 
and clusters and found a low mass slope of about -1 (B\"ohringer et al. 2017)
for the mass range covered by observations, 
$ \ge 3\times10^{12}h^{-1}_{70} M_\odot$. Fig. 3 shows that the slope of the  
numerical function decreases further below this limit to an asymptotic value of 
about 0.35 (i.e. $\alpha = - 1.35$ with ${dn\over dM} \propto  M^{\alpha}$).
This corresponds to the insignificant increase seen
in the mass fraction.

\section{Discussion and Conclusion}

We see in Fig. 1 that the matter density fraction in galaxy halos
and clusters match well at the low mass end, as well as the underlying
cumulative mass functions. The mass fraction of the
galaxy group and cluster fuction reaches
a saturation value of $\Omega_{\rm cluster virial}/\Omega_m=0.28(1\pm0.02)$
and the galaxy luminosity function leads to
$\Omega_{\rm galaxy virial}/\Omega_m=0.28(\pm0.08)$.
This provides a convergent answer for the mass contribution of 
collapsed objects if the region
considered is restricted to the pseudovirial radii. 
This means that the bulk of mass
is in the intergalactic space. We note that the result 
for galaxy halos does not change
when we use the values relevant to other colour bands.
With other colour band results (Blanton et al. 2001; McKay et al.
2002), we obtain the mass density
$u:g:r:i:z=0.64: 1.14: 1: 0.99: 1.03$, where we have normalised the
values to the $r$-band result. With the exception of the $u$-band,
which is strongly affected by star formation, we
have a convergent answer with variations well within the uncertainties, 
and we can take the value from the $r$-band as the representative 
mass of haloes within the pseudovirial radius of galaxies.

\begin{figure}
   \includegraphics[width=\columnwidth]{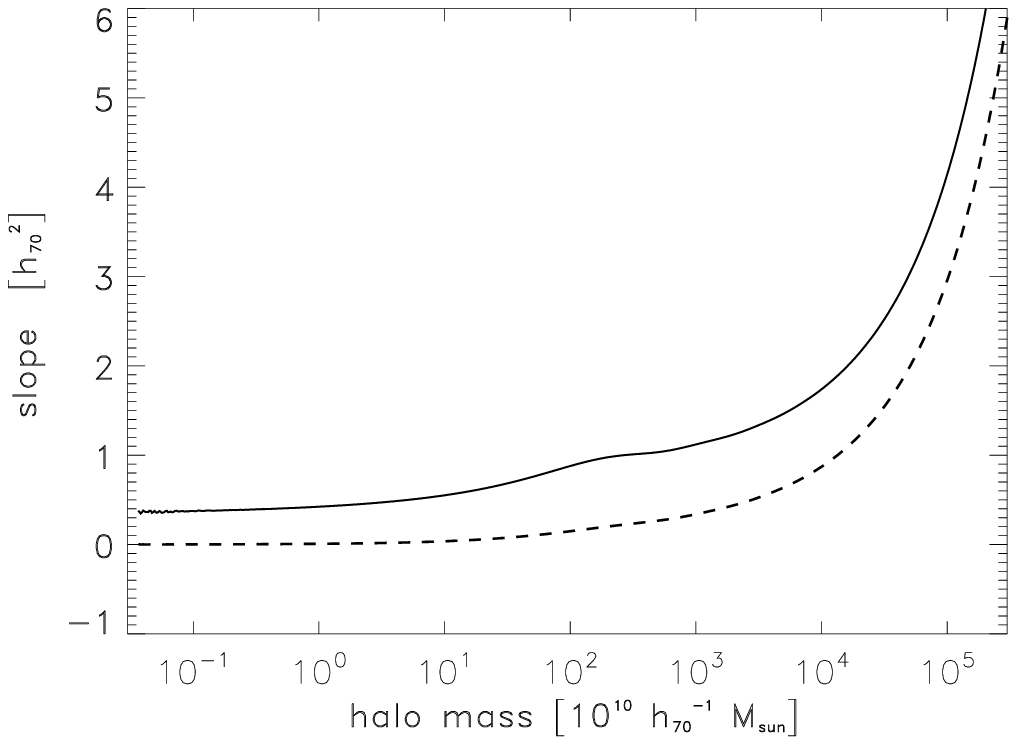}
\caption{Logarithmic slope of the cumulative mass function (solid line)
 and the matter density fraction (dashed line) of groups of clusters.
}\label{fig3}
\end{figure}

It is interesting to see that the cluster-group mass fraction 
function departs from the galaxy mass fraction 
for $M > 3 \times 10^{11}h_{70}^{-1}$,
indicating that cooling processes, which are essential 
for galaxy formation, become less effective for masses larger than this limit. 
This leads
to the observed high-mass cutoff of the mass function from galaxies, 
while the high mass cutoff for clusters and
groups is purely set by the intial condition and the gravitational
physics.

We see in Fig. 2 that most of the mass is contained in objects 
in the mass range $M_{200} \sim  10^{12} - 10^{14} h^{-1}_{70} M_\odot$.
It is worth noting, that this is the range of structures where
the variance of the density fluctuations in the linearly extrapolated
density fluctuation field, usually designated by $\sigma(M)$, is
close to unity. For the quoted mass range we find  
$\sigma(M) = 0.8 - 1.9$. Since we determine the
structure formation model that fits our observations best, we also derive
the variance of density fluctuations as a function of filter radius,
$\sigma(M(R_f))^2$. We
find $\sigma(M) = 1$ at $M_{200} \sim  5\times10^{13} h^{-1}_{70} M_\odot$.
This is the mass scale where most object formation takes place 
at the present epoch
and it is thus not surprising to find most matter in collapsed objects
in this mass range. 

The observations imply that substantially more mass is distributed
beyond the pseudovirial radius of $r_{200}$, for both galaxies and
clusters while it is custom to adopt $r_{200}$ to define the
cluster. The pseudovirial sphere contains only 28\% of the matter density
in the Universe.
This is in good agreement with the N-body result, which gives 26\% for the 
mass fraction contained within $r_{200}$ (MFY).
This increases to 45\%
within $2.2r_{200}$ and increase to 70\% if the radius of sphere is
taken to be 10 times $r_{200}$ (MFY).  Our results exhibit that
galaxies and clusters live at the peak patches of the density field,
and most of the mass is present in intergalactic space.  We stress
that this differs from the distribution of the luminous component (stars), which
should have an edge of the distribution, corresponding to the cooling
radius of the baryons. We expect that the hot gas behaves similarly to dark matter
at cosmological scales, where we see, at large radii, 
no reasons to segregate gas from
dark matter. So the fractions we discussed here are 
likely to apply similarly to the distribution of baryons.

\section*{Acknowledgements}
MF thanks late Yasuo Tanaka for the hospitality at the
Max-Planck-Institut f\"ur Extraterrestrische Physik and also Eiichiro
Komatsu at Max-Planck-Institut f\"ur Astrophysik in Garching, where the bulk
of this work was done. He also
wishes his thanks to Alexander von Humboldt Stiftung for the support
during his stay in Garching. He received in Tokyo a Grant-in-Aid
(No. 154300000110) from the Ministry of Education in Japan.
H.B. likes to thank Gyoung Chon for her role in the compilation 
and construction of the REFLEX data and for discussions.

%%%%%%%%%%%%%%%%%%%%%%%%%%%%%%%%%%%%%%%%%%%%%%%%%%



%\bibliographystyle{mnras}


%%%%%%%%%%%%%%%%%%%%%%%%%%%%%%%%%%%%%%%%%%%%%%%%%%

%%%%%%%%%%%%%%%%% APPENDICES %%%%%%%%%%%%%%%%%%%%%

%\appendix

%%%%%%%%%%%%%%%%%%%%%%%%%%%%%%%%%%%%%%%%%%%%%%%%%%

% Don't change these lines
\bsp	% typesetting comment
\label{lastpage}
\end{document}